\documentclass[nonacm]{acmart}

\AtBeginDocument{%
  \providecommand\BibTeX{{%
    \normalfont B\kern-0.5em{\scshape i\kern-0.25em b}\kern-0.8em\TeX}}}

\setcopyright{none}

 \acmConference[Unpublished working draft]{arXiV}
 \acmBooktitle{arXiV}

\begin{document}

\title{Look at Me When I Talk to You:\linebreak A Video Dataset to Enable Voice Assistants to Recognize Errors}

\author{Andrea Cuadra}
\email{apc75@cornell.edu}
\affiliation{%
  \institution{Cornell Tech}
  \country{USA}
}
\author{Hansol Lee}
\email{apc75@cornell.edu}
\affiliation{%
  \institution{Cornell University}
    \country{USA}
}
\author{Jason Cho}
\email{apc75@cornell.edu}
\affiliation{%
  \institution{Cornell University}
    \country{USA}
}
\author{Wendy Ju}
\email{wendyju@cornell.edu}
\affiliation{%
  \institution{Cornell Tech}
    \country{USA}
}

\renewcommand{\shortauthors}{Cuadra, et al.}

\begin{abstract}
People interacting with voice assistants are often frustrated by voice assistants' frequent errors and inability to respond to backchannel cues. We introduce an open-source video dataset of 21 participants' interactions with a voice assistant, and explore the possibility of using this dataset to enable automatic error recognition to inform self-repair. The dataset includes clipped and labeled videos of participants' faces during free-form interactions with the voice assistant from the smart speaker's perspective. To validate our dataset, we emulated a machine learning classifier by asking crowdsourced workers to recognize voice assistant errors from watching soundless video clips of participants' reactions. We found trends suggesting it is possible to determine the voice assistants's performance from a participant's facial reaction alone. This work posits elicited datasets of interactive responses as a key step towards improving error recognition for repair for voice assistants in a wide variety of applications.
\end{abstract}

\ccsdesc[500]{Human-centered computing}
\ccsdesc[500]{Computing methodologies~Intelligent agents}

\keywords{error-recognition, self-repair, voice assistants, conversational design}

\maketitle

\section{Introduction}

``Alexa, spell `Seven.' ''

\noindent ``Salmon is spelled S-A-L-M-O-N.''

Today's voice agents are blind to the giggling and the eyerolls of the people they are interacting with. On one hand, this may spare the digital feelings of machine assistants; on the other, it keeps the assistants from recognizing the errors they make and from learning from them. People often get frustrated with voice agents making frequent mistakes and not responding to any feedback other than explicit voice commands. 

In human face-to-face interaction, people monitor each other to see if their meaning is being understood by others, and they stop and self-correct if they recognize that they have made an error. Computers and machines could more easily perform error recovery if they used cameras to see how people react to their statements. For example, a machine would know it had just made an error if it could observe the user shaking his/her head in response to its actions. For machines to be able to make sense of these cues, however, they need some model of how people respond when they hear a correct versus an incorrect response.  

\begin{figure}[t]
\centering
\includegraphics[width=.85\columnwidth]{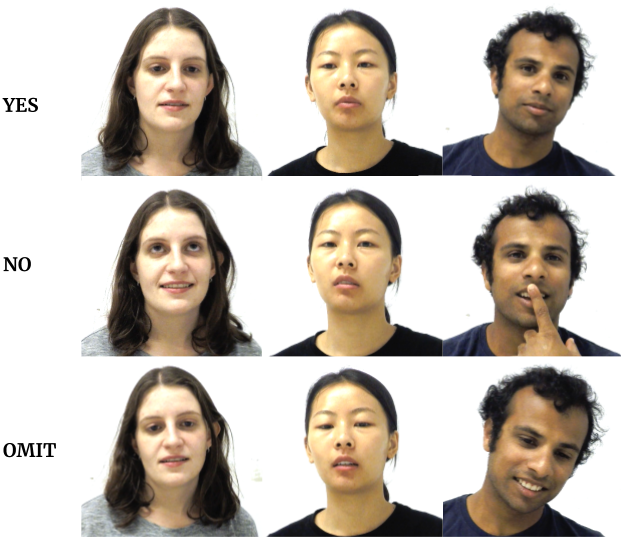}
\caption{Our dataset includes people reacting to correct responses (YES), incorrect responses (NO), or non-responses (OMIT).}
\vspace{-1em}
\label{fig:examples}
\end{figure}

In this work, we introduce an open-source video dataset of people reacting to a voice agent's right or wrong answers and explore how this dataset could be used to train machines to recognize error based on user's facial expressions by emulating a video classifier. This is a first step in a broader initiative to build a machine-learning model for voice assistants to recognize error based on people's reactions. For this project, we first collect video samples of people reacting to Amazon Alexa's right or wrong answers, then we remove the sound and ask crowdsourced workers to guess whether or not a mistake was made based on the participant's facial expressions. This work contributes on a novel approach in human machine interaction, using datasets of interactive responses to inform the development artificial intelligence to improve human machine interaction, and a dataset on which machine learning can be performed.
 
\section{Background}

Our dataset is constructed from people interacting with current-day voice agents, but the intent of the research is to inform the response people to have to voice assistants more generally. The advantage of working with current-day voice agents is that people are familiar with them, and their lack of embodiment presents a more generic class of responses than people might have in response to a speech-enabled robot or a screen-based avatar. Here, we survey a variety of related research from a number of arenas to make the case for the need for a more general model of how users respond to errors in speech recognition.

\subsection{Speech-enabled devices}
The commercial viability of voice assistants is presaged by the plethora of new voice-based devices and user interfaces now on the market. Many are voice-enabled assistants, such as Microsoft's Cortana \cite{MicrosoftCortana}, Amazon's Alexa \cite{AmazonAlexa},
Apple's Siri \cite{AppleSiri}, or Google Assistant \cite{GoogleAssistant}.
In some of the first-generation voice assistant research published in the larger HCI community, such as \cite{nielsen1992finding,bretan1995simulation, suhm2002comparative}, the computer-generated voice that people were speaking with was often on the other side of a phone line. In other cases, like the embodied conversational agents of Cassell, Sullivan, Churchill and Prevost \cite{cassell2000embodied}, the voice was front-ended by on-screen virtual agents.  Unlike in these initial examples, today's voice agents are embodied in standalone devices such as Amazon's Echo \cite{AmazonEcho} or Dot, Apple's Homepod \cite{AppleHomepod}, or Google's Home or Home mini \cite{GoogleHome, GoogleHomeMini}, and are usually situated proximate to people in their living or working quarters. 

The widespread explosion in speech-enabled interfaces is driven by advances in natural language processing, text-to-speech and dialog generation systems driven by big data, as well as hardware breakthroughs in far-field microphone arrays. 
While improvements to the hardware and software of these speech-enabled devices might improve the recognition of individual words people say, common-sense intelligence is not yet in grasp \cite{winograd_economist_2017}. The limited capabilities of today's speech systems would be well complimented by interactive savvy that would help them recognize and recover from conversational errors.

\subsection{Repairing Error}

Voice agents may make mistakes, but human dialog is far from error-free itself. A key difference is that people perform repair in communication \cite{schegloff1977preference}, monitoring listeners to see if they have been heard and understood before moving forward in the conversation. This process, in which speakers seek and provide evidence of understanding, is called grounding. \cite{clark1986referring} In collaborative conversations \cite{clark1989contributing}, addressees must therefore also indicate their understanding, or lack of understanding, to help the speaker understand the state of the communication. 

Linguists Schegloff, Jefferson and Sacks define repair to be the practices that interactants use to handle troubles in hearing, speaking and understanding that occur regularly in social interaction. \cite{schegloff1997practices, schegloff1997third, schegloff2000others}
They found from analyzing naturalistic conversation that people had a preference for self-correction over being corrected by others: in moments when repair was necessary or possible, the distribution of repairs was strongly skewed towards self-repair \cite{schegloff1977preference}. Very often self-repair occurs when the speaker notices a mistake, in the transition space between speaking turns, before the listener even has a chance to respond. 


 Gieselmann ran a small experiment to look at what error recovery strategies people use when talking to robots compared to when they talk to other people. Gieselmann found that very different repair tactics were used, largely due to the limited interaction capabilities of robot, and that the most common indicator that an error was made is a sudden change in the dialogue topic. In this research, the focus of the error detection lay in analysis of the discourse \cite{gieselmann2006comparing}. More recently,  Salazar-Gomez et al. experimented with using EEG-based feedback methods to correct robot mistakes in real time;  because the EEG signals were analyzed in real-time in closed-loop fashion, the robot was able to respond to possible signs of error by hyper-articulating actions to elicit stronger response to help it determine if it was making a mistake \cite{7989777}.

\subsection{Facial Displays for Conversational Facilitation}
Between human interactors, the visual feedback channel can play an important role in the coordination of joint understanding. Clark and Krych performed an experiment where one participant directed another on assembling Lego models; pairs which could not visually see each other performed much slower if the director could not see the builder's workspace, and much worse when the instructions were audio-taped. \cite{clark2004speaking} The face in particular is critical to providing feedback in discourse. Birdwhistell noted in 1970 that facial displays perform linguistic functions, particularly as listener commentaries. \cite{birdwhistell1970} Ekman and Friesen characterized the category of nonverbal acts which maintain and regulate the back-and-forth nature of speaking and listening as \textit{regulators}. \textit{Regulator} actions occur in the attentional periphery; people perform them without thought, but can recall and repeat them if asked. Addressees are sensitive to the lack of these cues, but are rarely aware of them when they are present. \cite{ekman1969repertoire} In 1991, Chovil performed an experiment with people listening to a story in a face-to-face, partition, and telephone and answering machine condition, and found that listeners primarily react facially when they would be seen by the storyteller \cite{chovil1991social}. 

The importance of identifying and incorporating responses to such conversational signals was recognized early in the human-computer interaction community by Nagao and Takeuchi \cite{nagao1994speech}. While linguists and behavioral psychologists have recognized and analyzed the \textit{regulatory} use of facial displays, the machine learning and computer vision community has largely focused on emotion recognition in their analysis of faces \cite{Paiva2017, Lubis2018}. This is in part due to the widespread availability of emotional expression image databases such as Ekman's Pictures of Facial Affect \cite{ekman1976pictures}, the Belfast database \cite{douglas2000new}, the Extended Cohn-Kanade Dataset \cite{lucey2010extended}, or the Affectiva-MIT Facial Expression Dataset \cite{mcduff2013affectiva}. These datasets have been used for a wide variety of applications, such as to evaluate effectiveness of advertisements \cite{teixeira2014and} or political branding \cite{mcduff2013measuring}. 

While emotion is certainly an important factor in interaction, strong shows of emotion often take place only after the user is angry, and the interaction is beyond repair. \cite{batliner2003find} Analysis of facial displays for conversational grounding and efficacy should be just as important, since linguists have found, at least anecdotally, that at least two-thirds of facial actions in dialogue are communicative rather than emotional expressions \cite{fridlund1987facial}. Bousmalis et al. for instance, have surveyed the conversation analysis literature for nonverbal audiovisual cues that indicate agreement and disagreement between human speakers, with the goal of developing machine recognition of these cues. \cite{Bousmalis}

\subsection{Embodiment in Conversational Facilitation}
The human-robot interaction community has also examined affect recognition \cite{breazeal1999build, rani2006empirical}, but, in that community, there is greater recognition of the use of embodied signals for conversational \textit{regulation}. Fujie et al. made a robot that recognized head motions, like nodding, for paralinguistic information that clarifies speaker intent \cite{fujie2004conversation}. Sidner et al. found that participants that knew their robots recognized conversational head nods would nod more. \cite{sidner2006effect} Huang and Mutlu have proposed developing a Robot Behavior Toolkit that uses the social cues that people use to achieve interaction goals to make robots that are able to adapt their behaviors to people. \cite{huang2013repertoire} Mutlu et al. have focused predominantly on gaze cues to signal attention and intent, largely for humanoid robots \cite{mutlu2006storytelling,mutlu2009footing,mutlu2012conversational,andrist2014conversational,huang2015using,pejsa2015gaze}. 

We know of no research in the DIS, HCI, or HRI communities to date that focuses on facial displays for error recognition. Part of the challenge for these communities has been the lack of datasets focusing on facial displays for conversational facilitation to train machine learning models. Our project seeks to address this shortcoming.

\begin{figure}[t]
\centering
\includegraphics[width=.4\columnwidth]{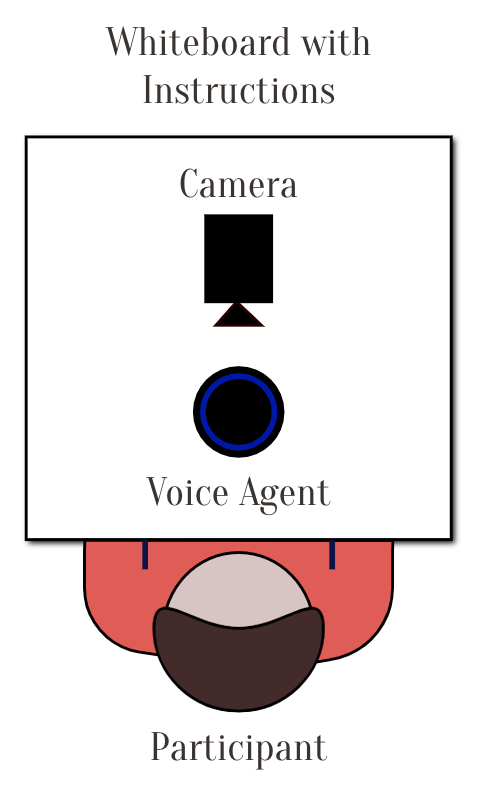}
\caption{Illustration of the setup: participants sat in front of the voice agent facing a camera and a whiteboard with handwritten instructions.}
\label{fig:setup}
\end{figure}

\section{Dataset Generation}

To generate the dataset of people reacting to voice agent answers, we recruited \textit{N} = 21 participants to come interact with an Amazon Tap with Alexa in a lab setting for approximately 30 minutes each. The participants were prompted to ask simple queries with right or wrong answers. They were given a set of example questions, but were instructed to ask anything they would like. Participants were compensated with \$5 gift cards each for their participation, and signed a consent form that informed them that ``Your participation will be audio- and video-recorded, will be used for academic \& public presentations, and made publicly available as part of an open video dataset.'' 

\subsection{Recording setup}
Fig. \ref{fig:setup} shows the setup of the room. Participants were video-recorded throughout the course of the the interaction, with the camera situated so that the video reflected the vantage point of the speech-enabled device. Aside from the instruction and training segment of the sessions, the actual recorded portion of the sessions was generally about 15 minutes long.

\subsection{Clip formation}
Participant recordings were manually trimmed and classified 
by a researcher into separate clips. The clips start when the voice agent starts talking, and end after the participant's main reaction to that response occurs, in most cases when the participant says "Alexa" to start a new request. 

\subsection{Yes/No/Omit}
We originally planned to compare only Yes and No responses, but found that Amazon Alexa missed participant commands often enough that an Omit category was necessary to address how people reacted when the voice agent failed to respond to their request at all. 

From the participant's perspective, the Omit and No conditions are similar, but from the machine perspective, they are completely different. In the case of \textit{Yes vs. No}, the machine knows it has just said something, and is evaluating whether it made an error. In the case of Omit, the machine does not know that any request has been made, and would need to distinguish the Omit reaction from the reaction of participants who are just idle.

\subsection{Resulting dataset}
\begin{figure}[t]
\centering
\includegraphics[width=1\columnwidth]{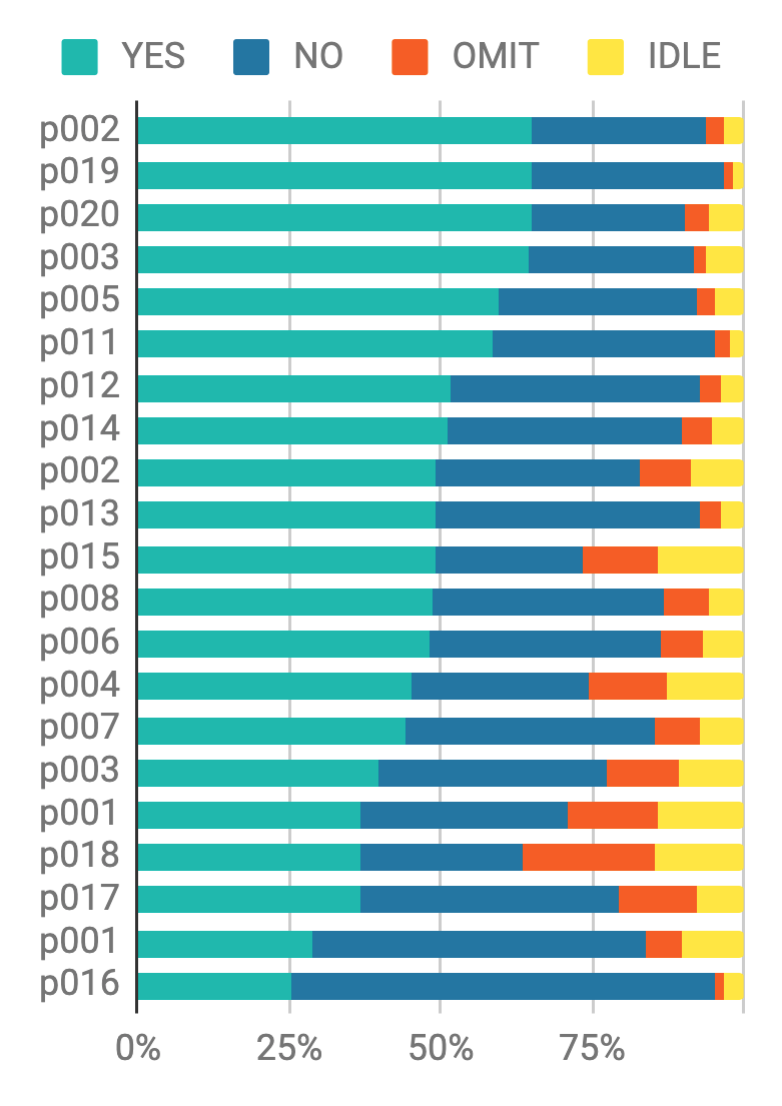}
\caption{Prevalence of labels by participant.}
\label{fig:preva}
\end{figure}

\begin{figure}[t]
\centering
\includegraphics[width=.5\columnwidth]{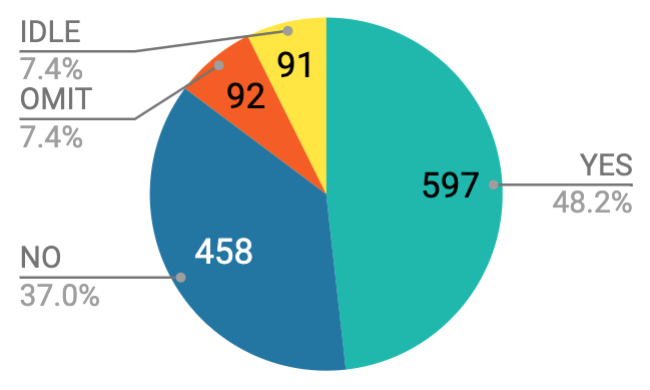}
\caption{ Out of 1147 reaction clips, Alexa answered correctly (YES) in 597, incorrectly (NO) in 458, and did not respond (OMIT) in 92. We additionally clipped 91 non-reaction (IDLE) clips.}
\label{fig:counts}
\end{figure}

The resulting dataset contains 1238 clips from 21 participants. Alexa answered correctly (YES) in 597, incorrectly (NO) in 458,  did not respond (OMIT) in 92 clips, and were idle (IDLE) in 91 clips. 
Figure \ref{fig:counts} illustrates the distribution of reactions in the dataset, Figure \ref{fig:preva} illustrates the distribution of reactions per participant, and Figure \ref{fig:examples} shows some examples of the types of reactions recorded. Note that in about half of all reactions, the voice agent is not telling the participant what the participant wants to hear.

To enable a point of comparison for the Omit condition, we created a set of $I$(Idle) clips in which participants were not waiting for Alexa to respond. We created approximately as many $I$(Idle) clips as $O$(Omit) clips. Clips vary in lengths from a couple of seconds to minutes. 

Each clip was labeled following this set of guidelines: 

$Y$ --- the voice agent successfully understood and responded to the participant; 

$N$ --- the voice agent misheard, misunderstood, was not able to fulfill, or misanswered the request; or

$O$ --- the voice agent did not hear or ignored the request.

$I$ --- the voice agent was idle, because no request was made.

Each clip file contains the reaction label ($N$, $Y$, $O$, or $I$), the location where the clip was recorded, the participant number, and a clip number. The enclosing directories also contain the dates in which the data was collected. Aside from the inclusion of participant faces, which are necessary given the nature of the study, the clips were anonymized. We had 21 participants, fourteen which identified as female, and the rest who identified as male. Eighteen were college students in their early twenties, and 3 were older adults. 11 participants were from the United States, and 10 participants were from outside the United States. 

\subsection{Dataset}

The dataset contains 1238 video clips of naturalistic reactions (or non-reactions) to voice agent interaction from 21 participants. The clips vary in length from a couple seconds to minutes. The video clips are in color, include sound, and are at least 1352 x 760 DPI in resolution.

We are hosting the dataset on our institution's digital repository. (Following publication), access is freely available to people who secure IRB approval from their own institution for secondary data analysis, indicating that they will use the data only for the research purposes intended. This provision is intended to protect human participants, who gave consent for their videos to be used for scientific research and machine learning, but not for other purposes.

\section{Dataset Initial Analysis}

To validate the dataset, we performed data classification using human crowd workers, online. We implemented two sets of analyses: \textit{Yes vs. No}, and \textit{Idle vs. Omit}. The goal of this data validation is to show that the dataset has signal that is detectable by an human classifier, and thus can more likely be used by machine learning scientists to computationally classify people's reactions. 
\setcounter{subsection}{0}
\subsection{Method}
For the validation, video clips from our dataset were shown to human classifiers viewing the videos online, to test whether facial expressions can aid in helping humans or machines recognize error.  We intentionally took worsened the quality and quantity of the data to match what we anticipated a commercial voice assistant would be able to process in real time, which at the moment does not usually include the context awareness from previous interactions or audio-prosodic features such as rhythm, tone, or intonation. To do so, we removed the sound from the clips in question, trimmed the video to leave out the participants' questions, and reduced the resolution to make the videos easily streamable. We envision machine learning scientists doing similar modifications to our dataset to test their individual hypotheses. The clips were reduced in size to 640x360 pixels, and clipped to be at most 12 seconds long.  

This validation helps to establish baseline feasibility of our method; the addition of sound, higher resolution or longer reactions should improve the recognition rates, as would the addition of additional participants to the dataset.

\subsubsection{Human Classification}

In the machine learning community, human classification is sometimes used as a proof-of-concept and benchmark for machine classification \cite{shenoy2008human,  amershi2014power, parikh2010role}. While there are debates about the quality of Amazon Mechanical Turk worker \footnote{https://www.mturk.com} annotations or classifications \cite{fort2011amazon}, nevertheless, seeing that people are able to make a classification using only the data that would be provided to a machine classifier is often used as proof that there is enough information in the data for classification to occur.

To ensure quality and consistency in our classification task, we recruited Mechanical Turk workers with a HIT (Human Intelligence Task) approval rate higher than 95\% who were located in the United States. 
We used online Qualtrics \footnote{https://www.qualtrics.com} surveys with embedded videos which Mechanical Turk workers were asked to classify. The overall assignment lasted about 12 minutes, and we  paid workers \$1.80 to complete the assignment. 
\subsubsection{Classifying \textit{Yes vs. No}}
Before starting the prediction tasks, each worker completed an introductory tutorial with video demonstrations for each label category. Each video demonstration had three parts. The first part showed a clip showing a full interaction of a participant with Alexa. The clip included the sound, the participant's question, and the participant's reaction. Then a black screen with white text indicated, ``In this example, the person got what they wanted. In the real tasks, you will only see the reaction without sound.'' This was followed by a clip like the one they would be asked to rate. Then the worker was presented with a survey question, ``Did the person get what they wanted?,'' and had the option to say ``yes'' or ``no.''  This process was repeated with a video demonstration of a \textit{No} video. We used the workers responses on these initial trial questions to identify which workers successfully understood the prompt of the assignment. 


After the introductory tutorial, each assignment contained 23 videos: 13 videos randomly selected from the pool of Y-labeled clips, and 10 videos randomly selected from the pool of N-labeled clips. The proportions were determined based on label prevalence in the dataset. After watching each video, humans were asked, ``Did the person get what they wanted?'' and had the options to say ``yes'' or ``no.''

170 Mechanical Turk workers completed the assignment, out of which 125 accurately responded to the two initial training videos. We removed 45 entries from workers who did not answer the training correctly. In total, we analyzed 2875 \textit{Yes} or \textit{No} predictions. Table \ref{tab:humanYN} summarizes our findings.


\begin{table}[tbh!]
\centering
\caption{Human classification of \textit{Yes vs. No}}
\label{tab:humanYN}
\begin{tabular}{c|c|c}
N = 2875          & Predicted: YES & Predicted: NO \\ \hline
Actual: YES & 843 & 782             \\
Actual: NO  & 583 & 667            \\
\end{tabular}
\end{table}

\subsubsection{Classifying Idle vs. Omit}
The protocol for the \textit{Idle vs. Omit} classification followed the one we used for classifying \textit{\textit{Yes vs. No}}. We recruited workers using the same guidelines, and had them complete an introductory tutorial with video demonstrations of participants in the Idle or Omit scenarios. 

The Idle or Omit assignment lasted about 6 minutes, and we  paid  participants \$0.90 to complete the assignment. After the introductory tutorial, each assignment contained 12 videos evenly selected at random from the pools of Idle- and Omit-labeled clips. After watching each video, workers were asked ``Was this person initially waiting for Alexa to respond?'' and had the options to say ``yes'', or ``no.''

57 Mechanical Turk workers completed the assignment, out of which 38 accurately responded to the two initial training videos. We removed entries from 19 workers who did not answer the training correctly. Like for the Yes/No Human Classifier, even though we removed responses in which people failed the training, we estimate that about 25\% of the workers that did not watch or understand the training videos might have accurately responded to the tutorial questions by chance as there were only two 2-option questions. We analyzed a total of 456 Idle or Omit predictions. Table \ref{tab:humanIO} summarizes our findings.

\begin{table}[tbh!]
\centering
\caption{Human classification of IDLE vs. OMIT}
\label{tab:humanIO}
\begin{tabular}{c|c|c}
N = 456         & Predicted: IDLE & Predicted: OMIT \\ \hline
Actual: IDLE & 129              & 99             \\
Actual: OMIT  & 111              & 117            
\end{tabular}
\end{table}

\section{Results}
The human classifiers performed slightly better than chance at predicting the labels of videos, 2.5\% better at distinguishing Y- and N-labeled clips, and 3.9\% better at distinguishing I- and O-labeled clips. The paired t-tests comparisons with chance \textit{Yes vs. No} and \textit{Idle vs. Omit} were, .0642 and .1128, respectively. Only the \textit{Idle vs. Omit} finding was statistically significant at a 90\% confident level. The \textit{Yes vs. No} comparisons suggest a trend that needs further testing.

\section{Discussion}

The process of classifying and validating the videos in our dataset turned out to have a number of complexities that were non-obvious, and which we feel would be good for anyone researchers following in our footsteps to know.

\subsection{Complexity in Error Classification}
One surprising aspect of this process was that the separation of YES/NO/OMIT clips was not straightforward, and involved a greater degree of subjectivity than the authors initially expected.


Coding the ground truth for each clip of whether an answer was right or wrong was challenging. For example, there are many ways in which Alexa can be wrong, and many ways in which a person might be displeased by Alexa's response. Only some of these cases should require Alexa to perform self-repair, but how might we distinguish a frown due to a bad odor in the room from a frown due to Alexa hearing the name of the wrong song? Additionally, some responses are layered, having correct and incorrect components --- for example, in one case, a German participant asks about the weather in Dortmund, Germany. When Alexa gets the location right, the person is satisfied; however, fractions of a second later, Alexa says the weather in degrees Fahrenheit, so the person immediately becomes dissatisfied as he was expecting the weather in degrees Celsius. 

Retrospectively, as we dug deeper into these issues, we discovered several works suggesting error taxonomies for chat and speech systems. Higashinaka et al. distinguished utterance-level, response-level, context-level and environment level errors, and further proposes multiple subcategories to each \cite{higashinaka2015towards}. Bohus and Rudnicky's work suggests the following non-understanding errors: out-of-application (conversational level), out-of-grammar (intent level), ASR Error (signal level) and End-pointer error (channel level) \cite{bohus2005sorry}. These taxonomies were helpful to us when classifying errors, and will be useful again when deciding what type of repair to perform. However, the distinction between errors did not seem to cause differences that we could detect in the human reaction.

\subsection{Can People Detect Error?}
This validation of our reaction dataset with human classifiers was performed to test whether we could recognize interaction errors using facial reactions alone. The results from using humans to recognize error based only on visual cues suggest that error recognition is possible, but that this is a very difficult task, even for people. Our human classifiers performed only slightly better than chance. This is an important finding on its own, and opens up ways in which our dataset can be used. For example, instead of making the clips low-resolution and silencing them, dataset users can keep the clips' original resolution and audio content. By doing so, they can do a multimodal study similar to what D'Mello and Graesser did over ten years ago \cite{d2010multimodal}. In a multimodal semi-automated affect detection study, D'Mello and Graesser found that the accuracy of a multichannel model (face, dialogue, and posture) was statistically higher than the best single-channel model for fixed (but not spontaneous) affect expressions \cite{d2010multimodal}. This suggests that combining the audio and visual elements from our dataset could yield surprisingly better results, and a dataset like ours opens up possibilities for many to do so without having to do they data collection step themselves. In retrospect, we should have done our initial validation analysis with more than just one modality, but doing another round of testing was beyond the scope of this work, especially given our findings were sufficient to demonstrate the dataset's usefulness.

On the whole, this validation indicates that facial reactions \textit{can} be used by machine learning scientists to further explore the possibilities for error-recognition to enable recovery. This dataset brings us a first step closer at enabling voice agents to successfully recognize error in order to perform self-repair to reliably improve one-on-one interactions. 


\subsection{The cost of repair}
One key question that our finding that people are only slightly better than chance at recognizing error from only facial displays uncovers is: how good does error recognition need to be for successful interaction? Is it possible that repair and recovery works as a process in human dialogue \textit{despite} our weak recognition of error because attempts at conversational repair are not costly? 

After improving error-recognition accuracy, the most immediate next step is to focus on self-repair. In the current work, the modeling of likely error occurred post-facto, and off-line. By developing models for error recognition that can occur on-line, we would be able to have machines experiment with different ways that a voice agent could perform self-repair. Bohus, et al. performed online supervised repair of error telephone-based spoken dialog system that provides bus route and schedule information in the greater Pittsburgh area \cite{bohus2006online}. Similar experiments for in-situ voice interaction could employ a similar approach to on-line learning; we expect the right repair policy probably differs in different contexts, and it would be interesting to see how visual approaches to error recognition and subsequent recovery differ from speech-based approaches. 

\subsection{Hedging}
Based on our qualitative observation of the interactions, the main point of confusion was often that Alexa states her answers very definitively, as if she were sure of her answers. When Alexa is making a mistake, the mismatch between that tone and the actuality of the error causes people to take longer to understand that a breakdown has occurred. This type of error, an error in belief or conviction of response, is one that Alexa makes with regularity that seems not to be named at all in aforementioned taxonomies \cite{higashinaka2015towards,bohus2005sorry}. 
If a respondent is giving an answer she is unsure about, she hedges by expressing some level of uncertainty in her voice and expressions. This provides ``feedforward'' information to the requester to provide more marked signals about the success or failure of the answer, and is usually accompanied by greater attention by the respondent to the requester's reactions of approval or further confusion. This type of closed loop interaction 
could be made possible if the computer's responses were modulated by its confidence that it had the right response to the user's query.


\subsection{Scaling the method}

To collect data at a larger scale, we believe it is important to come up with a method and setup that motivates longer streams of interaction between the user and the voice agent.  There are many ways to achieve this goal; for example, by using scenarios where there are many faces simultaneously providing input about the same event, like the one in Kateva's Robot Comedy Lab in which a robotic comedian collects facial data from a large number of people at once \cite{Katevas2015}. 

Also, it is possible that defining a task for participants to perform with the help of the voice agent might have yielded a larger dataset of interactions. In this study, although we made suggestions for topics for people to discuss with the voice agent, we left the actual queries up to users. We felt people's facial reactions were more likely to be natural if they were  motivated by real interest in the answers. In retrospect, however, this setup also had the effect of limiting the number of interactions, because people randomly recruited for our study did not intrinsically have a lot of things they really wanted to ask the voice agent. 

It could also be that the best way to deploy this experiment is not in a laboratory setting, but in situ, employing code in ``beta'' chatbots that collects reactions ``in the wild,'' and then asking users how they felt in order label the reactions. Learning individual models would also help improve error recognition accuracy as we noticed more consistency in reactions within participants, rather than between participants. 

\subsection{Priming}
The final reflection we had on our method is that we could have primed participants to believe that the machine was looking for visual cues in order to recognize its own errors. Previously cited research from Fujie et al. and Sidner et al. had found that participants who knew that robots were responding to their head nods would nod their heads more \cite{fujie2004conversation, sidner2006effect}; by priming the participants, we might help them to become more expressive and responsive when interacting with the voice agent. More pronounced visual signals could help increase prediction accuracy. Another experiment to measure the value of visual cues could compare error recognition in clips with only sound or only video against clips with sound and video. The combination of audio and visual cues might be a lot more useful than either set of cues alone.

Finally, it is also possible that people interacting with a machine that interactively performs self-repair would be more emotive, and more pointedly make facial displays that reflect positive or negative feedback, making the tasks of predicting easier.

\subsection{Ethical Considerations}
We recognize that both the collection of people's faces and the analyzing of their expressions, can introduce a wide set of ethical concerns, for example, about privacy infringements \cite{Choe:2011:LGH:2030112.2030118, 8048642}, about giving personal information to companies 
\cite{jacoby2018}, and even about exposing people to risk of deep fakes \cite{knight_2017} built with their likenesses. Here, we mention the ethical considerations and implications that we encountered in this work. 

\subsubsection{Privacy}
One of the challenges of our dataset and other computer-vision datasets is that they are based on videos of actual people. Our videos were of participants who had gone through consent review, were offered incentives that were in line with fair wage compensation, and had given permission for their images and likenesses to be used for a machine learning dataset. 

Because our dataset includes the voices and faces of our participants, obviously personal identifiable information \cite{kolata_2019}, we are restricting our distribution of the dataset to researchers who receive an internal review board (IRB) clearance in advance, acknowledging limitations on the use of the videos for machine learning and interaction research purposes. 
We acknowledge the privacy, security and trust issues that arise from cloud computing in general, and facial recognition in specific \cite{lau2018alexa, valentino-devries_singer_keller_krolik_2018, solon_2019}. 
We do not advocate (nor do we find practical) the large-scale transmission and collection of face image data from people's personal environments to companies with voice-enabled products. It is our intent that our research eventually be built into algorithms and methods that enable real-time error recognition and response on the local devices in people's pockets, cars and homes.

\subsubsection{Fairness and Transparency for Crowdwork}
As mentioned previously, we recruited Mechanical Turk workers from the United States, and paid crowdworkers who performed our original classification task ($YES$/$NO$) 1.80 USD, and our second classification task ($IDLE$/$OMIT$) 0.90 USD. Based on our pilot testing of the task, we estimated that the first task should take 12 minutes, and the second 6 minutes. The equivalent hourly wage for the work is 9.00 USD, above the US Federal minimum wage \cite{silberman2018responsible, hara2018data}. The actual mean time on tasks was 10.8 minutes for the first task and 5.2 minutes for the second task. We approved payment for all completed tasks, regardless of quality of response.

\section{Conclusion}
This work marks the beginning of a much larger body of research to be performed in error-recognition and self-repair for human-machine interaction. We gathered videos of humans interacting with a voice agent, created clips to include the participant's reaction, and labeled them by type of reaction. We have verified this method as a way of moving towards developing classifiers and predictors of errors, and offer our dataset for others would be interested in this foundational interaction problem.

Moving forward, based on this experience, we have identified complexities in error classification, hedging, repair, priming and scaling of method as key new directions for research in this domain. 

Finally, we aim to apply this research towards applications of people interacting with speech devices in naturalistic settings.  Although some voice agent applications can easily have a fixed camera position, like voice agents in cars, future speech-enabled agents might be appliances or mobile robots in people's living or working quarters without a consistent field of view of their user(s). It is important to explore how error-recognition and self-repair plays out in these contexts. 

Today, we interact with incredibly smart machines with no common sense about interaction. The ability to negotiate conversation--not only to model the words that people are saying correctly, but to know if the conversation is going well or not--is a skill that people do exceedingly well, and one that has long been out of reach for machines. The ability to use the multiple channels and the interactive strategies that people use to negotiate meaning can make interactions with voice agents less stilted and frustrating. This research moves us in the direction of being able to create more natural human-machine interactions.

\begin{acks}
This research was deemed to be IRB exempt by Cornell University. We obtained consent for collection and dissemination of audio-visual data for the purposes of machine learning from each participant. We would like to thank David Goedicke, Jofish Kaye, and Chinmayi Dixit for their considerable help and advice, and the Mozilla Foundation for their financial support of this project.
\end{acks}
\bibliographystyle{ACM-Reference-Format}
\bibliography{lamwitty}

\end{document}